\documentclass[aps,prb,twocolumn]{revtex4-1}
\usepackage{amsmath,amssymb,bm}
\usepackage[dvipdfmx]{graphicx}

\usepackage{txfonts}
\usepackage{amsmath}
\usepackage{bm}
\usepackage{comment}
\usepackage{here}
\usepackage{kantlipsum}
\usepackage{color}
\usepackage{xspace}

\newcommand{\alphaI}{{$\alpha$-(BEDT-TTF)$_2$I$_3$}\xspace}

\begin{document}

\title{
Tilted Dirac Cone Effect on Interlayer Magnetoresistance
in $\alpha$-(BEDT-TTF)$_2$I$_3$
}

\author{
Naoya Tajima
}
\affiliation{
Department of Physics, Toho University, Miyama 2-2-1, Funabashi-shi,
Chiba 274-8510, Japan
}
\author{
Takao Morinari
}
\affiliation{
Graduate School of Human and Environmental Studies, Kyoto
University, Kyoto 606-8501, Japan
}

\date{\today}

\begin{abstract}
We report the effect of Dirac cone tilting
on interlayer magnetoresistance
in \alphaI, which is a Dirac semimetal
under pressure.
Fitting of the experimental data by the theoretical formula
suggests that the system is close to a type-II Dirac semimetal.
\end{abstract}

\maketitle

The discovery of an unconventional half-integer quantum Hall 
effect in graphene\cite{Novoselov2005,Zhang2005} 
has stimulated intensive research
on massless Dirac fermion systems.
When a conduction band and a valence band touch
at a single point in energy--momentum space
with linear energy dispersion,
the system is called a massless Dirac fermion system.
The touching points appear as pairs
known as Dirac points.
Despite the Fermi velocity
of those Dirac fermions being much smaller than the speed of light,
they are described by the relativistic Dirac equation.
The half-integer quantum Hall effect is a consequence
of their unusual electronic structure.\cite{Geim2007}
If the Dirac points are at the Fermi energy,
the system is called a Dirac semimetal.
A number of Dirac fermion systems have been
discovered, including surface states of 
topological insulators.\cite{Hasan2010}
There is also intensive research 
on Weyl fermions,\cite{Wan2011} which are
a two-component analog of Dirac fermions.

In general, the energy dispersion of Dirac fermions,
which has a cone-like shape called a Dirac cone,
is tilted from the energy axis in 
energy--momentum space.
In Dirac or Weyl semimetals,
both electron and hole pockets
appear if the tilt is large enough.
Such a system is called a type-II Dirac or Weyl
semimetal,\cite{Soluyanov2015}
where Lorentz invariance is broken,
and physical properties
are very different from the usual Dirac or Weyl fermions,
which are called type I.

In this paper,
we report the effect of the Dirac cone tilt
on the interlayer magnetoresistance
in \alphaI, which is a Dirac semimetal
under pressure.\cite{Kobayashi2004,Katayama2006,Tajima2006,Tajima2007,Kajita2014}
We found that the tilt of the Dirac cone 
is very large and the system is close to
a type-II Dirac semimetal.

To investigate the tilt of the Dirac cone,
we may consider a single Dirac point,
though there are two Dirac points in \alphaI\
because the tilt of  the other Dirac cone is the same.
The Hamiltonian is described 
by the following $2 \times 2$ matrix:\cite{Kobayashi2007}
\begin{equation}
 {\mathcal H}\left( {{k_x},{k_y}} \right) = \left( {\begin{array}{*{20}{c}}
{v_0^x{k_x} + v_0^y{k_y}}&{{v_x}{k_x} - i{v_y}{k_y}}\\
{{v_x}{k_x} + i{v_y}{k_y}}&{v_0^x{k_x} + v_0^y{k_y}}
 \end{array}} \right),
\end{equation}
where we set $\hbar=1$ and $(k_x,k_y)$ is the wave vector in 
the two-dimensional Brillouin zone.
The anisotropy in the Fermi velocity is parameterized by
$\alpha  = \sqrt {{v_x}/{v_y}}$.
The vector $(v_0^x,v_0^y)$ is associated with the tilt of
the Dirac cone.
The angle between the $k_x$ axis and the tilt direction
is defined by
\begin{equation}
 \gamma  = {\tan ^{ - 1}}\left( 
			  {
			  \frac{{v_0^y/{v_y}}}{{v_0^x/{v_x}}}
			  } 
			 \right).
\end{equation}
The tilt of the Dirac cone is described by the following parameter:
\begin{equation}
\eta  = \sqrt {{{\left( {v_0^x/{v_x}} \right)}^2} + {{\left(
		    {v_0^y/{v_y}} \right)}^2}}.
\end{equation}
If $\eta < 1$, the system is a type-I Dirac semimetal.
If $\eta > 1$, the system is a type-II Dirac semimetal.

Under a magnetic field 
${\bm{B}} = B \left( {\sin \theta \cos \phi ,
\sin \theta \sin \phi ,\cos \theta } \right)$,
the interlayer magnetotransport
is governed by the zero-energy Landau level
because the Fermi energy is at the Dirac point.\cite{Osada2008,Tajima2009}
At zero temperature, the interlayer resistivity
is given by\cite{Osada2008,Morinari09}
\begin{equation}
\rho _{zz} = \frac{A}{{{B_0} + B\sin \theta 
\exp \left[ { - \frac{1}{2}
{{\left( 
a_c/\ell _z
\right)}^2}
I\left( \phi  \right){{\cot }^2}\theta } \right]}},
\label{eq_formula}
\end{equation}
where $A$ is a parameter inversely proportional
to the density of states and 
the square of the interlayer hopping\cite{Osada2008} and
$B_0$ is a parameter associated with
impurity scattering.
The value of this latter parameter is estimated
as $B_0 = 0.7 $ T from the magnetic field
dependence of the interlayer magnetoresistance.\cite{Tajima2009}
$\rho_{zz}$ depends on the azimuthal angle $\phi$ 
through the following function:\cite{Morinari09}
\begin{eqnarray}
 I\left( \phi  \right) 
&=& \lambda {\left( {\alpha \sin \phi \cos \gamma  
- \frac{1}{\alpha }\cos \phi \sin \gamma } \right)^2}\\
& & + \frac{1}{\lambda }{\left( {\alpha \sin \phi \sin \gamma  
 + \frac{1}{\alpha }\cos \phi \cos \gamma } \right)^2},
\end{eqnarray}
where $\lambda=\sqrt{1-\eta^2}$,
$a_c$ is the lattice constant for the $c$ axis, and
${\ell _z} = 1/\sqrt {eB\sin \theta }$
is the magnetic length with $e$ being the electron charge.

From the analysis of the tight-binding model
for \alphaI,\cite{Kobayashi2007}
we find $\gamma=89.0^\circ$,
$\alpha=1.2$, and $\lambda=0.40$.
(For the other Dirac cone, we find $\gamma = 269.0^\circ$.)
When $\alpha \neq 1$, there is a contribution from the Fermi 
surface anisotropy to the $\phi$ dependence of $\rho_{zz}$.
However, $\alpha$ is close to 1, and so we 
set $\alpha=1.2$ in the following analysis.
As a result, the fitting parameters are
$A$, $\lambda$, and $\gamma$.

Experiments were conducted as follows: 
A sample on which four electrical leads 
(gold wire with a diameter of 15 $\mu$m) 
are attached by carbon paste 
was placed into a Teflon capsule filled with the pressure medium
(Idemitsu DN-oil 7373). The capsule was then set into 
a clamp-type pressure cell made of 
hard alloy MP35N, 
and hydrostatic pressure of up to 1.7 GPa was applied. 
The pressure was examined by recording the change in the
resistance of Manganin wire at room temperature. 
The resistance of the crystal was measured by using a conventional 
dc method with an electrical current of 0.1 $\mu$A 
along the $c$ crystal axis, which is normal to the 
two-dimensional plane. 
In the investigation, the interlayer magnetoresistance 
$\rho_{zz}$ was measured as a function of the azimuthal 
angle $\phi$ in a magnetic field of 7 T at 4.2 K.

\begin{figure}[htbp]
   \centering
     \includegraphics[width=0.8 \linewidth]{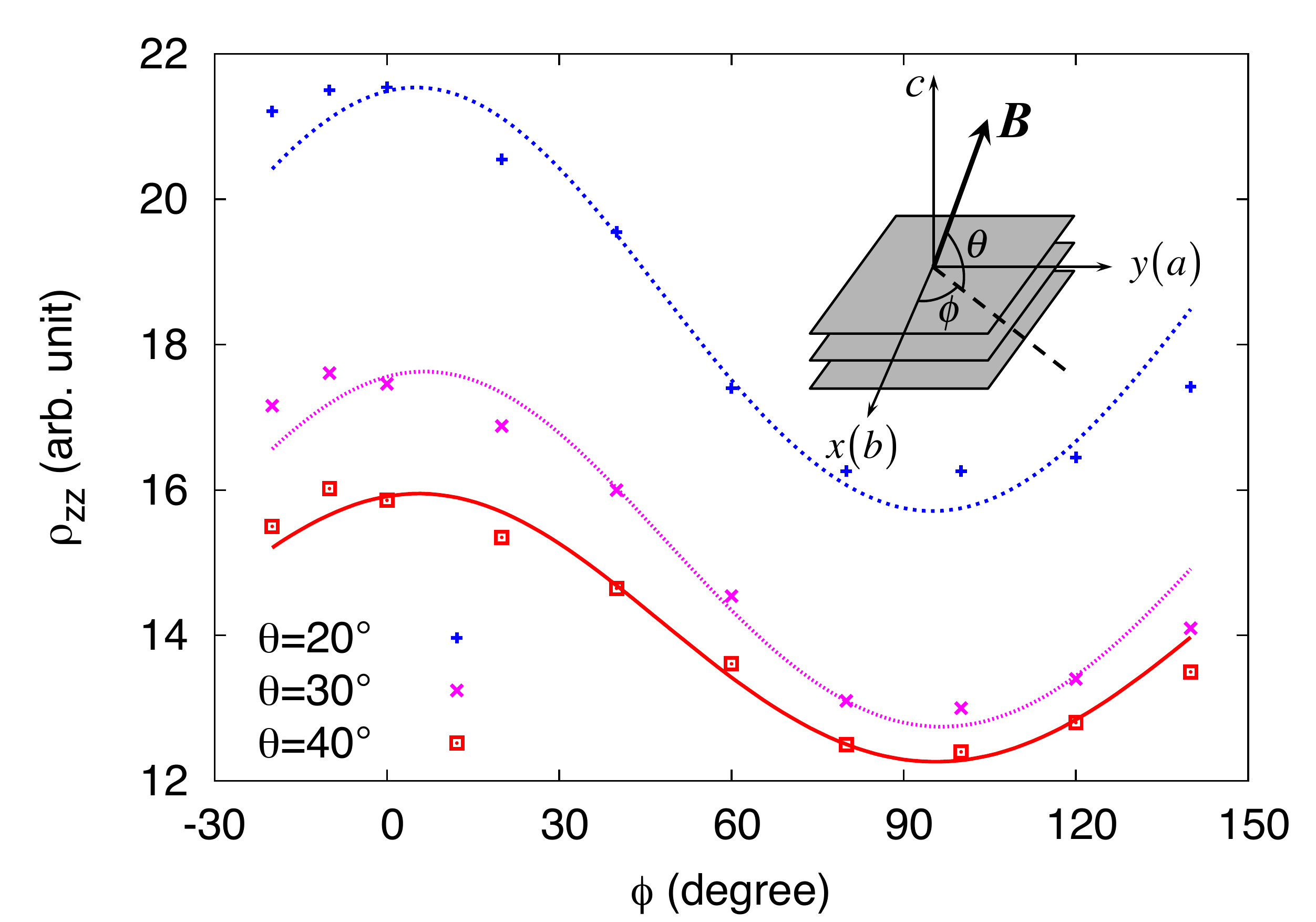}
    \caption{ \label{fig_result}
     (Color online)
   Azimuthal angle $\phi$ dependence 
   of the interlayer resistivity $\rho_{zz}$
   for different values of $\theta$.
   The experimental data are fitted by 
   using the theoretical formula (\ref{eq_formula}),
   which are shown by lines.
   The inset shows the layers of Dirac fermions and
   the crystal axes $a$, $b$, and $c$.
   The $b$ ($a$) axis corresponds to the $x$ ($y$) axis.
   }
  \end{figure}

The experimental result was fitted by formula (\ref{eq_formula}),
as shown in Fig.~\ref{fig_result}.
The parameter values obtained by the fitting
are listed in Table~\ref{table_res}.
From this analysis, we find that
$\eta$ is less than one but very close to one.
Therefore, the Dirac cone in \alphaI\
is almost at the boundary between
types I and II.
We also find that the direction of the tilt is approximately
along the $k_x$ axis, or the $b$ axis,
because $\rho_{zz}$ is maximum when the magnetic field
is in the direction of the tilt.\cite{Morinari09}
This is consistent with 
the tight-binding calculation\cite{Katayama2006}
and the first-principles calculation.\cite{Kino2006}
Note that the value of $\lambda$ increases
as $\theta$ decreases.
This behavior is understood as follows:
As $\theta$ decreases, mixing between the Landau levels
increases.\cite{Morinari10a}
This suppresses the anisotropy associated with the interlayer
hopping of the zero-energy Landau level wave function.
Meanwhile, $I\left( \phi  \right) \sim {\cos ^2}\phi /\lambda$,
for $\lambda \ll 1$.
Therefore, to describe the suppression
of the anisotropy using formula (\ref{eq_formula}),
we need a large $\lambda$ value.
By contrast, the parameter $\gamma$ does not depend
on $\theta$ because the Landau level mixing
does not affect the anisotropy.
In addition, the value is not much different from the
$\gamma = 1.0^\circ$ value evaluated from the tight-binding model.
The parameter $A$ increases as $\theta$ increases.
This is understood from the reduction of the density of states
at the Fermi energy owing to lifting of spin degeneracy
by the Zeeman energy.
\begin{table}[thbp]
 \centering
 \caption{
 Values of the parameters
 determined by the fitting shown in Fig.~\ref{fig_result}.
 }
 \label{table_res}
 \begin{tabular*}{0.95 \linewidth}{@{\extracolsep{\fill}}ccccc} \hline
  $\theta$ (degrees) & A (arbitrary units) & $\lambda$ & $\gamma$ (degrees) & $1-\eta$ \\ \hline
  40 & 55.9 & 0.0275 & 3.86 & $3.78 \times 10^{-4}$ \\
  30 & 47.1 & 0.0345 & 4.56 & $5.95 \times 10^{-4}$ \\
  20 & 43.1 & 0.0547 & 3.50 & $1.50 \times 10^{-3}$ \\ \hline
 \end{tabular*}
\end{table}

To conclude, we have measured the anisotropy
of the interlayer resistivity
and fitted the experimental data by using a theoretical formula.
From the analysis, we have found that
the Dirac cone of \alphaI\
is almost at the boundary between types I and II.
The signature of massive carriers 
in the in-plane mangetotransport\cite{Monteverde2013}
might be related to this fact.
Because the electronic structure of \alphaI\
is controlled by pressure,
we may expect that a type-II Dirac semimetal
is realized in \alphaI\ under high pressure, a subject
that is left for future research.

{\it Acknowledgments:}  This work was supported 
 by Grants-in-Aid for Scientific Research 
(A) (No. 15H02108),
(S) (No. 16H06346), and
(B) (No. 25287089) 
from the Ministry of Education, 
Culture, Sports, Science, and Technology, Japan.

\bibliographystyle{apsrev4-1}
\bibliography{../../../../work_space2/refs/tm_lib201801}

\end{document}